\documentclass[linenumbers]{aastex63}

\begin{document}
\shorttitle{JAGB (Milky Way) Zero Point}
\shortauthors{Madore et al.}
\graphicspath{{./}{figures/}}
%\begin{document}
\title{\bf Milky Way Zero-Point Calibration of the JAGB Method: \\ Using  Thermally Pulsing AGB Stars in Galactic Open Clusters} 
\begin{abstract} 
We present a new calibration of the J-band absolute magnitude of the JAGB method  based on thermally pulsing AGB stars that are members of Milky Way open clusters, having distances and reddenings, independently compiled and published by Marigo et al (2022). 17 of these photometrically-selected J-Branch AGB stars give $M_J = -6.40$~mag with a scatter of $\pm$0.40~mag, and a sigma on the mean of $\pm$0.10~mag. Combining the Milky Way field carbon star calibration of Lee et al. (2021) with this determination gives a weighted average of $M_J(MW) =$ -6.19 $\pm$ 0.04~mag (error on the mean). This value is statistically indistinguishable from the value determined for this population of distance indicators in the LMC and SMC, giving further evidence that JAGB stars are extremely reliable distance indicators of high luminosity and universal applicability. Combining the zero points for JAGB stars in these three systems, a value of $M_J = $ -6.20 $\pm$ 0.01 (stat) $\pm$ 0.04 (sys)~mag becomes our best current estimate of the JAGB zero point and its associated errors. Finally, we note that no evidence is found for any statistically significant dependence of this zero point on metallicity.
\end{abstract}
\keywords{Unified Astronomy Thesaurus concepts: Galaxy distances (590); Carbon Stars (199); Asymptotic giant branch stars (2100)}

\author[0000-0003-3431-9135]{\bf Barry F. Madore} 
\affil{The Observatories, Carnegie
Institution for Science, 813 Santa Barbara St., 
Pasadena, CA ~~91101, USA}
\affil{Department of Astronomy \& Astrophysics, University of Chicago, 5640 South Ellis Avenue, Chicago, IL 60637, USA}
\email{barry.f.madore@gmail.com} 

\author[0000-0003-3431-9135]{\bf Wendy~L.~Freedman}
\affil{Department of Astronomy \& Astrophysics, University of Chicago, 5640 South Ellis Avenue, Chicago, IL 60637, USA}
\affil{Kavli Institute for Cosmological Physics, University of Chicago,  5640 S. Ellis Ave., Chicago, IL 60637, USA}
\email{wfreedman@uchicago.edu}

\author[0000-0002-1576-1676]{\bf Abigail~J.~Lee}
\affil{Department of Astronomy \& Astrophysics, University of Chicago, 5640 South Ellis Avenue, Chicago, IL 60637, USA}

 \affil{Kavli Institute for Cosmological Physics, University of Chicago,  5640 S. Ellis Ave., Chicago, IL 60637, USA}
\email{abby@uchicago.edu}

\author[0000-0003-3339-8820]{\bf Kayla Owens}
\affil{Department of Astronomy \& Astrophysics, University of Chicago, 5640 South Ellis Avenue, Chicago, IL 60637, USA}
\email{kowens@uchicago.edu}

\keywords{Unified Astronomy Thesaurus concepts: Observational cosmology (1146); Galaxy distances (590); Carbon Stars (199); Asymptotic giant branch stars (2100);  Hubble constant (758)}
\section{Introduction}
The J-Branch stars of Weinberg \& Nikolaev (2021) have recently attracted renewed and wide-spread attention (Freedman \& Madore 2020, Madore \& Freedman 2020, Parada et al. 2021, Ripoche et al. 2020, Zgirski et al. 2021) as quality distance indicators of cosmological significance, especially given their very high luminosities at near-infrared wavelengths and the recent launch of JWST, which is fully capable of exploiting these stars as distance indicators well into the nearby Hubble flow. Moreover, these stars are easily and unambiguously identified by their very red colors in single-epoch exposures using filter pairs (equivalent to ground-based J \& K) that are part of 
JWST's imaging capabilities using NIRCAM and NIRSS.
In the broader context of their application to cosmology and the extragalactic distance scale in specific, JAGB  stars are capable of seriously competing with both the Cepheids and with the TRGB method. They are a magnitude brighter than the TRGB stars, even in the NIR. They are ubiquitous, being found in all Hubble types of galaxies (from ellipticals to lenticulars to spirals and irregulars), unlike Cepheids which are found only in star-forming galaxies. They are populous in number, with over 8,000 of them cataloged to date in the LMC, which is a dwarf galaxy. And they have a moderate dispersion in the absolute J-band luminosity of only $\pm$0.35~mag, which for a sample of only 400 stars would give an error on the mean distance modulus of only 0.018 mag, which converts to a statistical error on the mean of less than one percent in distance. \vfill\eject
\begin{figure*} 
\centering 
\includegraphics[width=18.0cm, angle=-90]{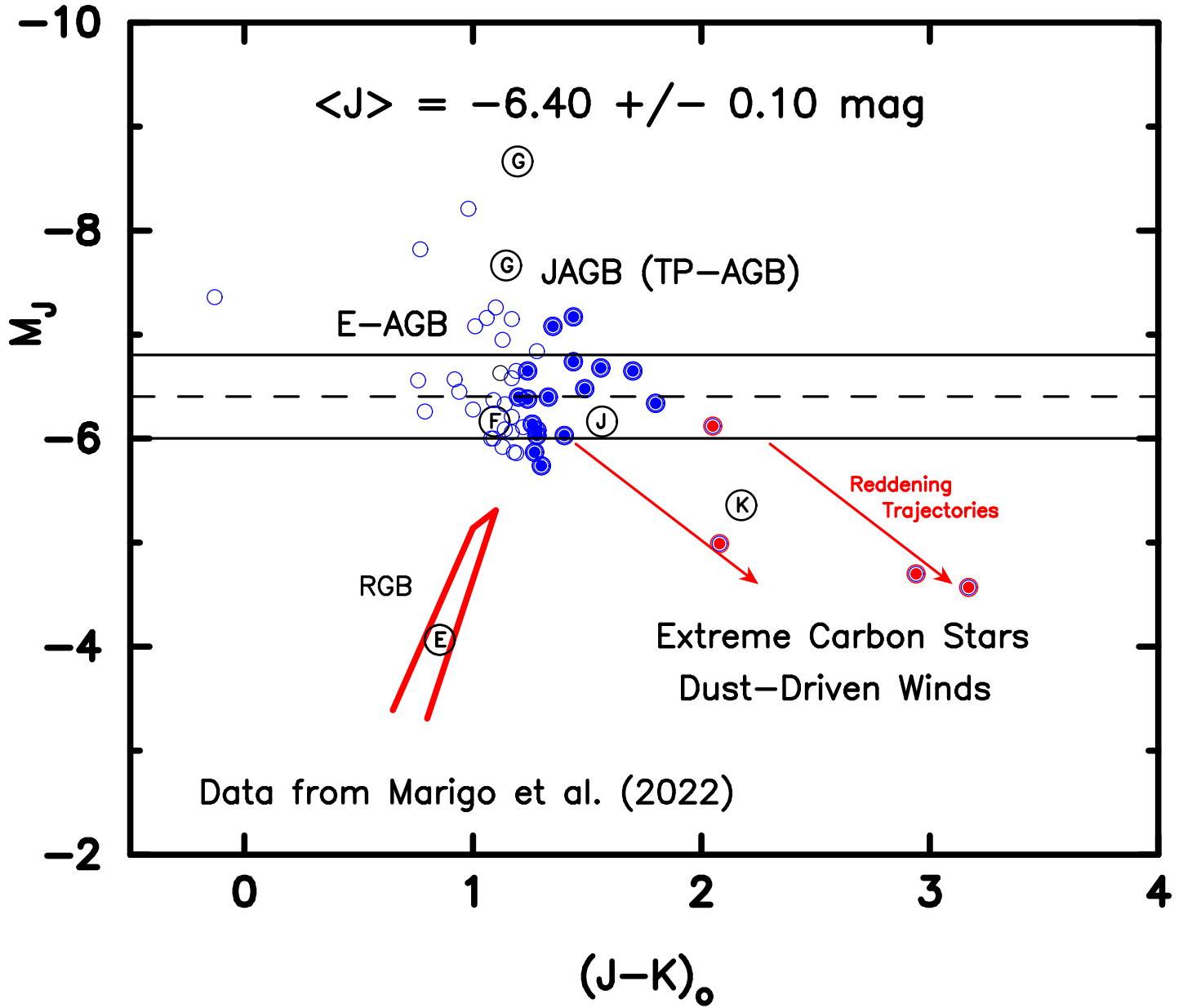} 
\caption{The NIR J vs (J-K) Color-Magnitude Diagram for AGB stars in Milky Way open clusters, as given in Table 1 of Marigo et al. (2022). Blue circled points (TP-AGB stars, according to Marigo et al.) are almost all within the JAGB region (WN01); open circles are her ``Early" Asymptotic Giant Branch (E-AGB) stars, and are photometrically within the Oxygen-rich AGB plume called ``Region G" (WN01). The circled letters (E, G, J \& K) each correspond to the same lettered regions given in Figure 1 of WN01.  
Red arrows bounding ``Region K'' are reddening trajectories in the J-(J-K) plane.}
\end{figure*}

\section{AGB Stars in Milky Way Open Clusters}
The NIR J vs (J-K) Color-Magnitude Diagram for AGB stars in Milky Way open clusters is given in Figure 1. These data are taken from Table 1 of Marigo et al. (2022), who gathered the distances and reddenings to the clusters directly from the catalog of Cantat-Gaudin et al. (2020). It is immediately noteworthy that the stars classified by Marigo as being TP-AGB (Thermally-Pulsing Asymptotic Giant Branch) stars (shown in blue) are photometrically identical to the color-and-magnitude selected JAGB stars that were first independently defined by Weinberg \& Nikolaev (2001, WN01 hereafter) as ``Region J''. Four ``extreme carbon stars'' are seen in this plot, falling below and to the red of the JAGB region, in what WN01 called ``Region E'', which is now thought to be populated by highly self-reddened carbon stars, blowing dust-driven winds. The open circles are what Marigo calls E-AGB stars, which photometrically belong to the bluer Oxygen-rich class of AGB stars ascending the ``I Branch'' (again, in the nomenclature of WN01). The red giant branch (WN01:``E Region''), with its upward sloping tip in the NIR, is shown in its red outline, slightly to the blue and about one magnitude fainter than the JAGB region. The black horizontal dashed line marks the mean magnitude of the JAGB stars at $M_J = -6.40$~mag. Its one-sigma bounds are shown by two black line flanking the mean magnitude by $\pm$0.35~mag. WN01 found an identical dispersion of $\pm$0.40~mag for their Table 1 sample of 8,229 JAGB stars in the LMC.

Ultimately, then, the utility of JAGB stars in establishing an independent and credible cosmological distance scale of their own will depend on the calibration of their zero point using as many independent geometrical determinations as possible, and additionally in making all of the necessary tests for hidden systematics along the way. The first geometric zero point was provided by the detached eclipsing binary (DEB) distance determined by Pietrzynski et al. (2019) for the LMC, giving $M_J (LMC) = $ -6.22 $\pm$0.01 (stat)~mag as used by Madore \& Freedman (2020) and Freedman \& Madore (2020). The second determination used JAGB stars in the SMC which also has a DEB distance (Graczyk et al. 2020), resulting in $M_J (SMC) = $ -6.18 $\pm$0.01 (stat)~mag. Next up were 153 field carbon stars in the Milky Way for which Gaia EDR3 parallaxes were available, giving $M_J = $ -6.14 $\pm$0.05 (stat)~mag (Lee et al. 2021). Finally, a value of -6.20 ~mag was confirmed systematically at the +0.02~mag level, by (Freedman \& Madore 2020) in an inter-comparison of JAGB distances to 14 nearby galaxies also having TRGB distances. At the time of writing, HST observations (PI: Hoyt) have been partially completed for several thousand JAGB stars in an outer field of the relatively-nearby galaxy, NGC~4258. This galaxy has a geometric distance (Reid et al. 2019) derived from the combined radial velocities and proper motions of water mega-masers orbiting the central black hole in this galaxy to derive a geometric distance. With this short {\it Letter} we return to the Milky Way to independently derive a JAGB zero point calibration in the NIR, using carbon stars known to be members of galactic open clusters with reddenings and distances for their main sequence population of stars given in the catalog of Cantat-Gaudin et al. (2020). This catalog was generated using a machine learning algorithm to identify clusters from the Gaia DR2 parallaxes. They determined individual distances and reddenings to each cluster based on isochrone fitting of the Gaia color-magnitude diagrams. Using these distances thus avoids the known systematics of the Gaia parallaxes (e.g., Lindegren et al. 2018, 2021).

%\vfill\eject
\section{A Brief Comparison with Theory}
It is generally agreed (see Marigo et al. 2008) that the maximum luminosity attained by AGB stars in their final stages of evolution are a monotonic function of their masses. In the sample studied here Marigo et al. (2022)  have provided main-sequence turn-off masses for the progenitors of each of the AGB stars in Figure 1. 
In the absence of significant mass loss between the main sequence and the AGB phase it is fair to assume that we know at very least the mass ordering of the AGB stars in the phase where their evolution is horizontal, which is precisely where the JAGB stars are positioned. 
 
Figure 2 shows the absolute J-band magnitudes of the 49 AGB stars in Table 1 of Marigo et al. (2022) plotted as a function of the main-sequence turn-off mass of the open cluster in which the AGB star has been found to be a member.The E-AGB stars, that generally have masses greater than 3.33$M_{\sun}$, are shown as open circles. The Extreme Carbon stars (thought to have dust-driven winds) are shown as red circled dots. And the thermally-pulsing AGB stars are shown as blue circled points in the mass range 1.16 to 3.33$M_{\sun}$. We further highlight the JAGB (sub-population of TP-AGB stars) by doubly-circling those 17 blue points. 
The remaining TP-AGB stars are to the blue of the JAGB stars in the CMD, and are still ascending the oxygen-rich AGB sequence, which has a random mix of ages at any given luminosity, as expected. 
\begin{figure*} 
%\centering 
\includegraphics[width=16.0cm, angle=-90]{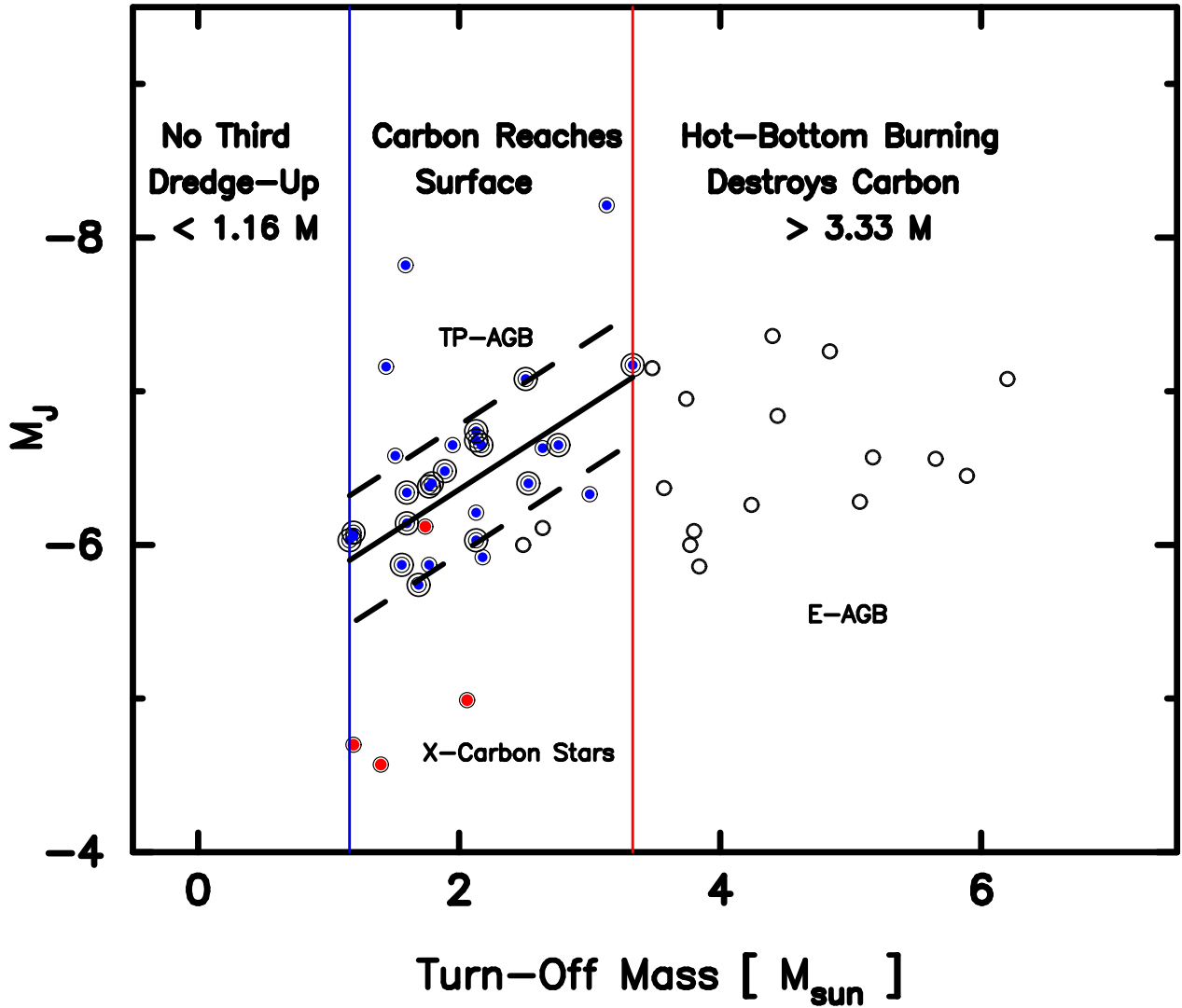} 
\caption{The absolute J-band magnitude of each AGB star found in an open cluster as a function of the main sequence turn-off mass of the host open cluster.  TP-AGB stars are shown as circled blue dots. E-AGB stars are black circles, and 
extreme (dusty) carbon stars are shown as circled red dots. The 17 color-selected JAGB stars from Figure 1 are shown as doubly-circled blue dots. 
The increase in $M(J)$ with turn-off mass for the JAGB stars, between M(TO) = 1.16 and 3.33 $M_{\sun}$, is shown by the solid line of slope -0.55 ($\pm$0.12) mag/$M_{\sun}$, which is flanked by 2-sigma (broken) lines, where $\sigma = 0.21$~mag.}
\end{figure*}

The lowest-mass TP-AGB star (as classified as Marigo et al.) is found at a turn-off mass of 1.16 $M_{\sun}$ (marked by a blue vertical line); while the highest-mass JAGB star is seen at 3.33 $M_{\sun}$ (delimited by a red vertical line). Higher-mass stars (black circled dots, as in Figure 1) range up to more than 6 $M_{\sun}$, and are all thought to be excluded from becoming carbon stars 
by the onset of their mass-dependent ``hot-bottom burning'' which destroys the newly synthesized carbon before it can be drawn to the surface. No stars in the Marigo tabulation have masses less than
1.16 $M_{\sun}$, below which it is thought that the third dredge-up (responsible for the first episode of bringing carbon to the surface) does not occur. No third dredge-up, at one extreme, and the onset of hot-bottom burning at the other, act to throttle the mass range in which AGB stars are capable of becoming a carbon stars, and, in doing so, this also restricts the luminosity range of these JAGB stars: making them standard candles.

That said, there is some tentatively identified structure within the JAGB population, as seen in this particular projection of this multi-dimensional data set. The upward-sloping black line, tracing the majority of JAGB stars through this diagram as a function of mass, is significant. This theoretically-predicted trend has an empirical least-squares slope of -0.55 $\pm$ 0.12~mag/$M_{\sun}$. To our knowledge, this is the first direct observation evidence that the absolute magnitudes of individual JAGB stars are themselves a function of their main-sequence progenitor's mass, and if significant mass loss has not occurred then this trend is suggesting that the observed J magnitude is also a function of the the star's current stellar mass.

The fact that the above mass-luminosity relation is not dispersionless is probably due to a combination of photometric errors (which could be improved) mixed with the possibility of small amounts of differential reddening (which will be harder to disentangle), and perhaps variable amounts of mass loss for stars of the same luminosity (which would represent an irreducible floor to the scatter seen here). 
The positive side of having identified the signal (predicted by theory) of increasing carbon-star luminosity with increasing AGB mass, means that the star formation history (SFH) of the galaxy will be imprinting a measurable/correctable signature of its own, reflected in fine structure in the JAGB luminosity function.
Once better understood and calibrated these features would allow for even higher-precision determinations
of the distance derived from this method. Even without sufficient numbers of carbon stars it may still be possible to predict these SFH corrections
using the more populated Oxygen-Rich (AGB) luminosity function which is the immediate precursor population of stars also reflecting the same SFH and mass range that is responsible for funneling select AGB stars into the JAGB region.

\medskip
\vfill\eject
\section{Summary Conclusions}
Shown in Figure 3, we now have two independently determined values for the J-band zero point of the JAGB population of carbon stars in the Milky Way: the above value
of $M_J = $ -6.40 $\pm$ 0.10 (stat)~mag determined from a select sample of 17 TP-AGB stars in open clusters, and the value obtained by Lee et al. (2021) using an independent set of 153 spectroscopically-selected Milky Way field carbon stars
using Gaia EDR3 parallaxes, giving $M_J = $ -6.14 $\pm$ 0.05~(stat) mag. Combining these two measurement, weighted inversely by their respective variances,
we arrive at a mean Milky Way value of  $M_J = $ -6.19 $\pm$ 0.04 (stat)~mag.

The value of the J-band zero point of the JAGB method derived above is for a Milky Way sample that is expected to have relatively high-metallicity progenitors. This can be compared to the zero-point determination made for stars in the LMC whose metallicities are generally lower than the Milky Way. And these two values can in turn be compared to the value for the SMC JAGB population which is thought to be at the lowest end of the metallicity range considered here. Those values are -6.19$\pm$ 0.04~mag (MW), -6.22 $\pm$0.01~mag (LMC) and -6.18~mag $\pm$0.01~mag (SMC), respectively. Without speculating as to what the exact metallicities are for each of these carbon-star samples, it is clear, from the small absolute-magnitude differences between them, and the lack of any monotonic trend (from high to low metallicity), that there is little evidence for any statistically-significant correlation of the mean J-band magnitude with metallicity, as sampled by this modest, but still representative sample of galaxies and their JAGB populations. The inverse-variance-weighted average of the three estimates is $M_J = $ -6.20 $\pm$ 0.01 (stat) $\pm$ 0.04 (sys)\footnote{Where the systematic error is the quoted uncertainty on the Detached Eclipsing Binary star method which was used to establish geometric distances to the LMC and SMC, (Pietrzynski, et al. 2019 and Graczyk et al. 2020), respectively}~mag; and this becomes our best current estimate for the JAGB zero point and its associated errors.
\begin{figure*} 
\includegraphics[width=12.0cm, angle=-90]{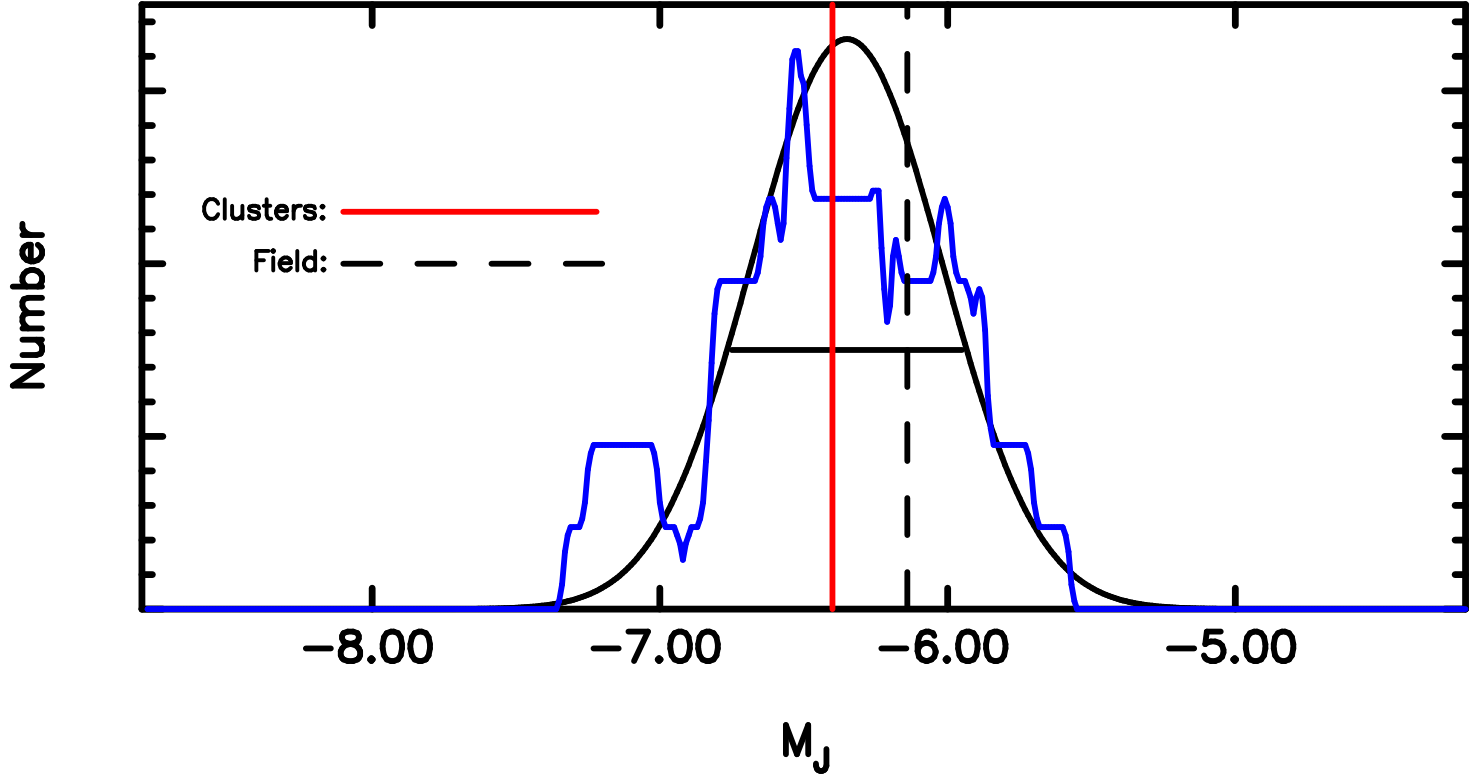} 
\caption{The J-band luminosity function (blue line) for the JAGB stars (blue points) seen in Figure 1. The mean absolute magnitude ($M_{\sun} = $ -6.40~mag) is shown by the vertical red line. The bold, dashed black vertical line to the right is the absolute J-band mean magnitude for the MW carbon stars individually calibrated by Lee et al.(2021) using Gaia EDR3 parallaxes. The black curve is a Gaussian with a sigma of $\pm$0.35~mag and a modal magnitude of -6.35~mag, fit to the centrally-located data between -7.0 and -5.5~mag only.}
\end{figure*}
\vfill\eject
\section{Acknowledgements}  

We thank the {\it University of Chicago} 
and {\it Observatories of the Carnegie Institution for Science} 
and the {\it University of Chicago} for their support of our long-term research into building up a secure calibration of the local distance scale and thereby incrementally allowing a better understanding of the expansion rate of the Universe and the systematics currently confounding those determinations. Support for this work was 
also provided in part by NASA through grant number HST-GO-13691.003-A 
from the Space Telescope Science Institute, which is operated 
by AURA, Inc., under NASA contract NAS~5-26555. This research made use of the NASA/IPAC Extragalactic Database (NED), 
which is operated by the Jet Propulsion Laboratory,
California Institute of Technology, under contract with the 
National Aeronautics and Space Administration. The critical reading of the paper by the referee is gratefully acknowledged.  Finally, we thank Dr. Paola Marigo for making the results of her research into AGB stars so easily accessible to the community.
\section{References}

\par
\noindent
Cantat-Gaudin, T., Anders, F., Castro-Ginard, A., et al. 2020, \aap, 640, 1

% \par
% \noindent
% Freedman, W.L. 2021, ApJ, 919, 16 arXiv: 2106.15656

\par
\noindent
Freedman, W.L., \& Madore, B.F. 2020, ApJ, 899, 67 

\par\noindent
Graczyk, D.,  Pietrzyński, I., Thompson, I., et al. 2020,  ApJ, 904, 13

\par
\noindent
Lee, A. J., Freedman, W. L., Madore, B. F., et al. 2021, ApJ, 907, 112.

\par
\noindent
Lindegren, L., Hern{\'a}ndez, J., Bombrun, A., et al. 2018, \aap, 616, A2

\par
\noindent
Lindegren, L., Bastian, U., Biermann, M., et al. 2021, \aap, 649, A4

\par
\noindent
Madore, B.F., \& Freedman, W.L. 2020, ApJ, 899, 66 

 \par
 \noindent
 Marigo, P., Girardi, L., Bressan, A., et al. 2008, A\&A, 482, 883
% \par
% \noindent
% Marigo, P., Girardi, L., Bressan, A., et al. 2017, ApJ, 835, 77.
\par
\noindent
Marigo, P., Bossini, D., Trabucchi. M., et al.  2022, ApJS, 258, 43
\par
\noindent
Parada, J., Heyl, J., Richer, H., et al. 2021, \mnras, 501, 933
\par
\noindent
Pietrzynski, G., Graczyk, D., Gallenne, A., et al. 2019, Nature, 567, 200
\par
\noindent
Ripoche, P., Heyl, J., Parada, J., \& Richer, H. 2020, \mnras, 495, 2858
\par
\noindent
Weinberg, M. D., \& Nikolaev, S. 2001, ApJ, 548, 712
\par
\noindent
Zgirski, B., Pietrynski, G., Gieren, W., et al. 2021, ApJ, 916, 19 

\vfill\eject
\end{document}